\documentclass[11pt,twoside]{article}
\usepackage{asp2010}

\resetcounters

\begin{document}

\title{Finding compact hot subdwarf binaries in the Galactic disc}
\author{T.~Kupfer$^1$, S.~Geier$^{2,3}$, A.~Faye McLeod$^{1,4}$, P.~J.~Groot$^1$, K.~Verbeek$^1$, V.~Schaffenroth$^{3,5}$, U.~Heber$^{3}$, C.~Heuser$^{3}$, E.~Ziegerer$^{3}$, R.~{\O}stensen$^4$, P.~Nemeth$^4$, V.~S.~Dhillon$^{6}$, T.~Butterley$^{7}$, S.~P.~Littlefair$^{6}$, R.~W.~Wilson$^{7}$, J.~H.~Telting$^{8}$, A.~Shporer$^{9,10,11}$ and B.~J.~Fulton$^{12}$}
\affil{$^1$Department of Astrophysics/IMAPP, Radboud University Nijmegen, P.O. Box 9010, 6500 GL Nijmegen, The Netherlands}
\affil{$^2$European Southern Observatory, Karl-Schwarzschild-Str. 2, 85748 Garching, Germany}
\affil{$^3$Dr. Karl Remeis-Observatory \& ECAP, Astronomical Institute, Friedrich-Alexander University Erlangen-N\"urnberg, Sternwartstr.
7, D 96049 Bamberg, Germany}
\affil{$^4$Institute of Astronomy, K.U.Leuven, Celestijnenlaan 200D, B-3001 Heverlee, Belgium}
\affil{$^5$Institute for Astro- and Particle Physics, University of Innsbruck, Technikerstr. 25/8, 6020 Innsbruck, Austria}
\affil{$^6$Department of Physics and Astronomy, University of Sheffield, Sheffield S3 7RH, UK}
\affil{$^7$Centre for Advanced Instrumentation, University of Durham, South Road, Durham DH1 3LE, UK}
\affil{$^8$Nordic Optical Telescope, Apartado 474, 38700 Santa Cruz de La Palma, Spain}
\affil{$^9$Las Cumbres Observatory Global Telescope Network, 6740 Cortona Drive, Suite 102, Santa Barbara, CA 93117, USA}
\affil{$^{10}$Department of Physics, Broida Hall, University of California, Santa Barbara, CA 93106, USA}
\affil{$^{11}$Division of Geological and Planetary Sciences, California Institute of Technology, Pasadena, CA 91125, USA}
\affil{$^{12}$Institute for Astronomy, University of Hawaii, Honolulu, HI 96822, USA}

\begin{abstract}
We started a new project which aims to find compact hot subdwarf binaries at low Galactic latitudes. Targets are selected from several photometric surveys and a spectroscopic follow-up campaign to find radial velocity variations on timescales as short as tens of minutes has been started. Once radial variations are detected phase-resolved spectroscopy is obtained to measure the radial velocity curve and the mass function of the system. The observing strategy is described and the discovery of two short period hot subdwarf binaries is presented. UVEX\,J032855.25+503529.8 contains a hot subdwarf B star (sdB) orbited by a cool M-dwarf in a P=0.11017\,days orbit. The lightcurve shows a strong reflection effect but no eclipses are visible. HS\,1741+2133 is a short period (P=0.20\,days) sdB most likely with a white dwarf (WD) companion.\\
\end{abstract}

\section{Introduction}
About half of the known single-lined hot subdwarf B stars (sdBs) are short period binaries with periods between a few hours to a few days. Binary evolution models show that the observed periods can be explained by a common envelope phase with a spiral-in of the companion during the red-giant phase of the sdB progenitor \citep{han02, han03}. Many studies have been performed to determine the orbital parameters of sdB binaries and reveal the nature of the companion stars. However, most of the systems studied so far are located at high Galactic latitudes (e.g. Morales-Rueda et al. 2003; Geier et al. 2011; Copperwheat et al. 2011\nocite{mor03, gei11, cop11}).\\
In the course of the MUCHFUSS project, \citet{gei13a} discovered the sdB binary with the shortest known period (P=0.048979\,days), CD-30\,11223 (see also Vennes et al. 2012; Heber et al. 2013\nocite{ven12, heb13}). In a follow-up work \citet{gei13b} showed that this system is a good candidate to explode as a type Ia supernova. They also showed that the progenitor system is a member of the Galactic disc with masses between 2-3\,M$_\odot$. Therefore, CD-30\,11223 belongs to a young stellar population rather than to the old population of stars at high Galactic latitudes. This means that there might be a different population of hot subdwarf binaries hidden at the largely unexplored low Galactic latitudes. \\
Inspired by this discovery, we started a campaign to search for short-period hot subdwarf binaries at low Galactic latitudes with periods of a few hours. Targets were selected from the UV-excess survey of the Northern Galactic plane (UVEX; Groot et al. 2009\nocite{gro09}), from a sample of bright sdBs identified using UV-colors from the Galaxy Evolution Explorer (GALEX) all-sky survey, and from the sdB database \citep{ost06}.\\ 
Here, we report on the observing strategy and present the discovery of the first two short-period sdB binaries, UVEX\,J032855.25+503529.8 (UVEX\,J0328) and\\ HS\,1741+2133 (HS\,1741).\\ 

\subsection{The UVEX and the GALEX sample}\label{sec:sample}
The UVEX survey images the northern Galactic plane (--5$^\circ<b<$+5$^\circ$) in the optical $U$, $g^{\prime}$ and $r^{\prime}$, and partly the narrowband He\,{\sc i} 5875 filters from $g^{\prime}$=14\,mag down to $g^{\prime}$=21\,mag to select UV-excess sources such as e.g. hot subdwarfs and white dwarfs \citep{gro09}. In a pilot study \citet{ver12} conducted spectroscopic follow-up of candidate hot stars discovered in the course of the UVEX survey. As expected, the majority of the objects are DA white dwarfs. However, \citet{ver12} showed that hot subdwarfs can be reasonably well separated from the bulk of DAs in a combination of colour and brightness. The contamination of DA white dwarfs is very low for magnitudes brighter than $g^{\prime}<$17\,mag. Most of the DA white dwarfs have (He\,{\sc i}--$r^{\prime}$)$<$-0.05 as they show no helium absorption, whereas the peculiar non-DA white dwarfs have (He\,{\sc i}--$r^{\prime}$)$>$--0.05 because of the helium absorption line. \\
The UVEX footprint spans 1850\,deg$^2$. Up to now about 1250\,deg$^2$ have been observed and photometric classification is available for 211\,deg$^2$. From the 211\,deg$^2$ about 70 UVEX objects have a $g^{\prime}<$17\,mag. Fig.\,2 in \citet{ver12} shows that about 40\% of their spectroscopically identified targets brighter than 17$^{th}$\,mag are either sdBs or hot subdwarf O stars (sdOs). That means we expect to find about 30 hot subdwarfs brighter than 17$^{th}$\,mag in the first 211\,deg$^2$ and therefore in the whole UVEX sample about 245 hot subdwarfs brighter than 17$^{th}$\,mag.\\
The GALEX sample consists of about 100 sdBs and sdOs at low Galactic latitudes (--30$^\circ<b<$+30$^\circ$). All targets are bright with magnitudes $B<$15 and selected from different sources, either from the subdwarf database \citep{ost06}, from spectroscopically identified hot subdwarfs, or from colour selected GALEX targets \citep{ven11,nem12}.\\
That means that the whole sample, once UVEX is completely available, is expected to contain about 350 hot subdwarf stars at low Galactic latitudes.

\begin{figure}[t]
\centering
   \begin{minipage}[b]{0.49\textwidth}
 \includegraphics[width=\textwidth]{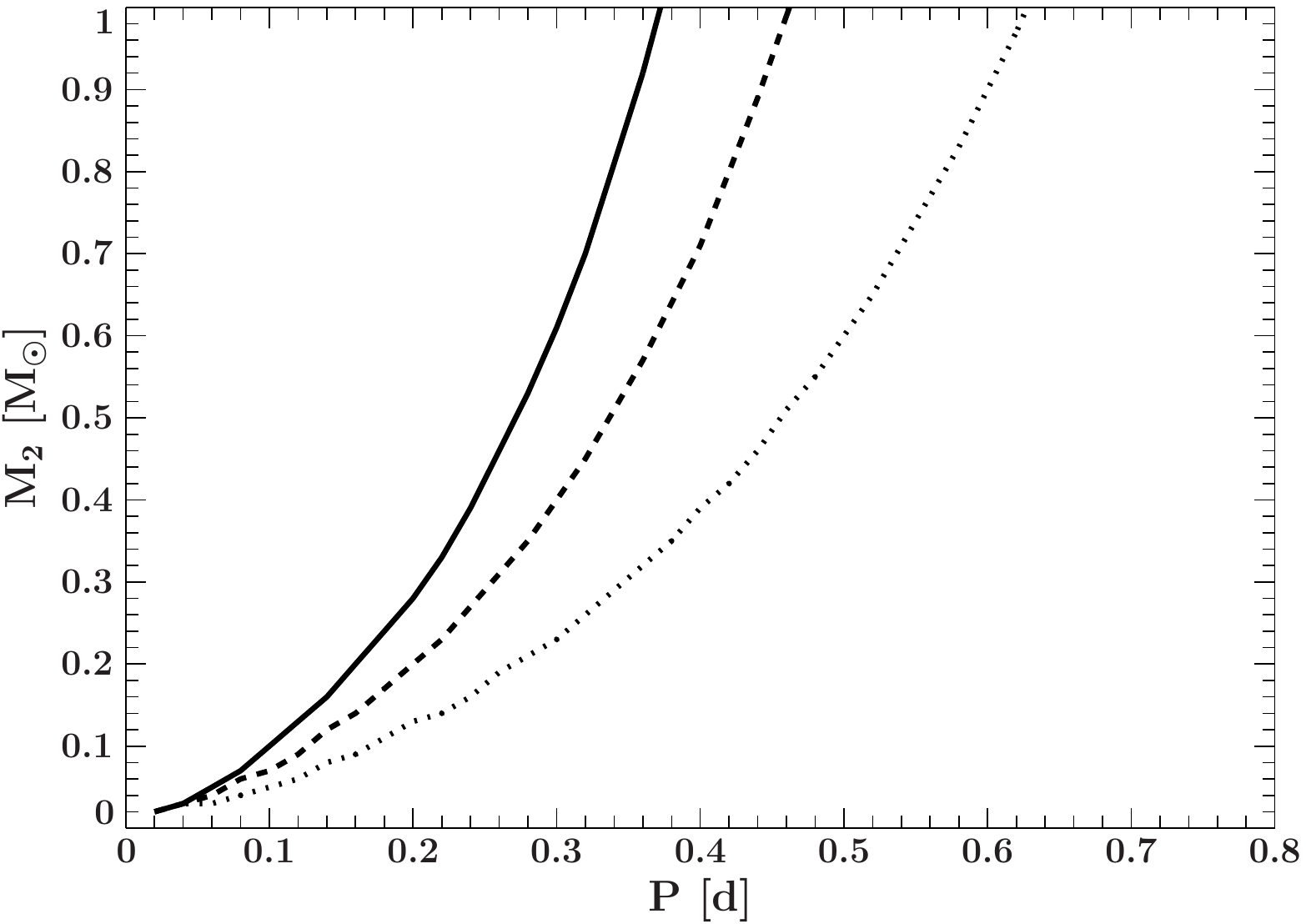}
\end{minipage}
    \hspace*{0.05cm}
       \begin{minipage}[b]{0.49\textwidth}
 \includegraphics[width=\textwidth]{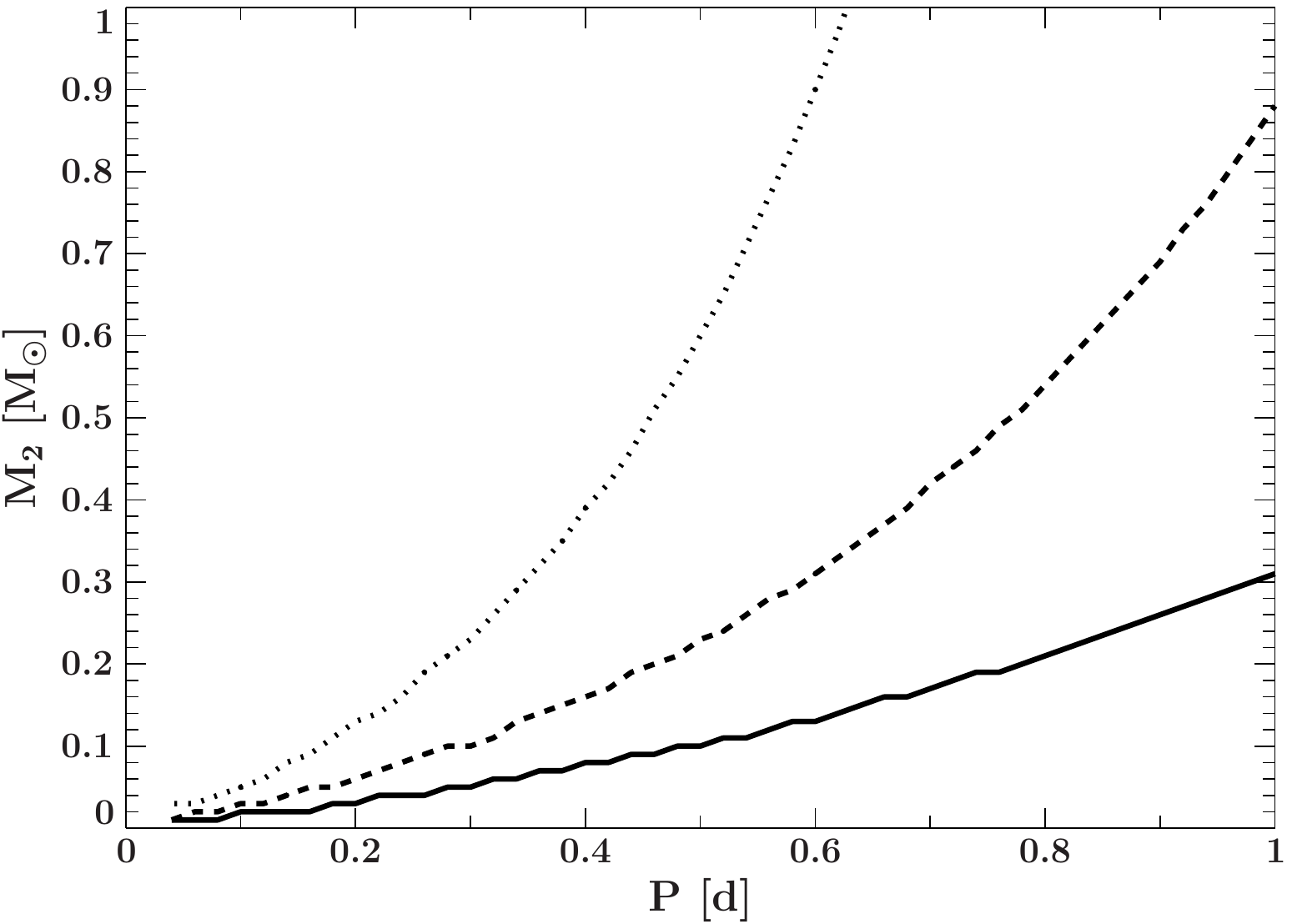}
\end{minipage}
\caption{Detectable companion mass for a given period assuming a primary mass of 0.47\,M$_\odot$, an inclination angle of $i$=90$^\circ$ and the system was observed around the zero-crossing in the RV curve. {\bf The Left hand panel} assumes different sequence observing times $\Delta t$=15\,min (solid line), 20\,min (dashed line) and 30\,min (dotted line) and a conservative uncertainty on the RV measurement $v_{\rm err}$=20\,km\,s$^{-1}$. {\bf The Right hand panel} assumes different uncertainties for the RV measurement $v_{\rm err}$=5\,km\,s$^{-1}$ (solid line), 10\,km\,s$^{-1}$ (dashed line) and 20\,km\,s$^{-1}$ (dotted line) and a sequence observing time of $\Delta t$=30\,km\,s$^{-1}$.} 

\label{fig:PvsM2}
\end{figure}

\subsection{Observing strategy and sensitivity for binary detection}
The goal of this project is to find hot subdwarf binaries with short periods and hence radial velocity (RV) variations on timescales on the order of tens of minutes. Therefore, a sequence of three spectra per object taken back-to-back with an arc after the first exposure is taken. The radial velocities are measured by fitting a set of mathematical functions (Gaussians, Lorentzians and polynomials) to the hydrogen Balmer lines as well as helium lines if present using the FITSB2 routine \citep{nap04}. If a significant shift between the three spectra is detected phase-resolved spectroscopy will be performed to measure the radial velocity curve and consequently the mass function of the system is derived. \\
For spectroscopy we use either the Isaac Newton Telescope using the IDS spectrograph ($R$=$\lambda$/$\Delta\lambda \simeq$1400, $\lambda$=3850 - 5360\,\AA) and the William Herschel Telescope using the ISIS spectrograph ($R$=$\lambda$/$\Delta\lambda\simeq$4000, $\lambda$=3650 - 5260\,\AA).\\
We ran tests to derive the sensitivity of our observing strategy. To estimate the expected shift for a given system we calculated a set of RV curves with different orbital periods $P$ and companion masses $M_2$, under the assumption of a primary mass $M_1$=0.47\,M$_\odot$ and an inclination angle of $i$=90$^\circ$. As simplification we assume that the system was observed around the zero-crossing in the RV curve (i.e. when the RV curve goes from blue to red or from red to blue). From the calculated RV curves we calculate the expected shift $\Delta v$ in the time interval $\Delta t$=10 - 30\,min around the zero-crossing because the observing time for a sequence of three spectra will be about 10 - 30\,min depending on the exposure time for an individual spectrum. We estimate an uncertainty for the RV measurement to be $v_{\rm err}$=5 - 20\,km\,s$^{-1}$ which will depend on the signal-to-noise ratio (SNR) in each spectrum. A detection is identified if the shift $\Delta v$ is larger than two times the uncertainty for the RV measurement $v_{\rm err}$ within a sequence of spectra $\Delta t$. Fig.~\ref{fig:PvsM2} shows that we should be able to detect typical WD companions up to periods of several hours. Low mass companions with masses of about 0.1\,M$_\odot$ are detectable in systems with periods of about 0.1\,days as well.\\
Simulations and first results (see Sec.~\ref{sec:results}) show that variations of $\sim$20\,km\,s$^{-1}$ between two subsequent spectrum can be detected.

\begin{figure}[t]
\centering
 \includegraphics[width=0.5\textwidth]{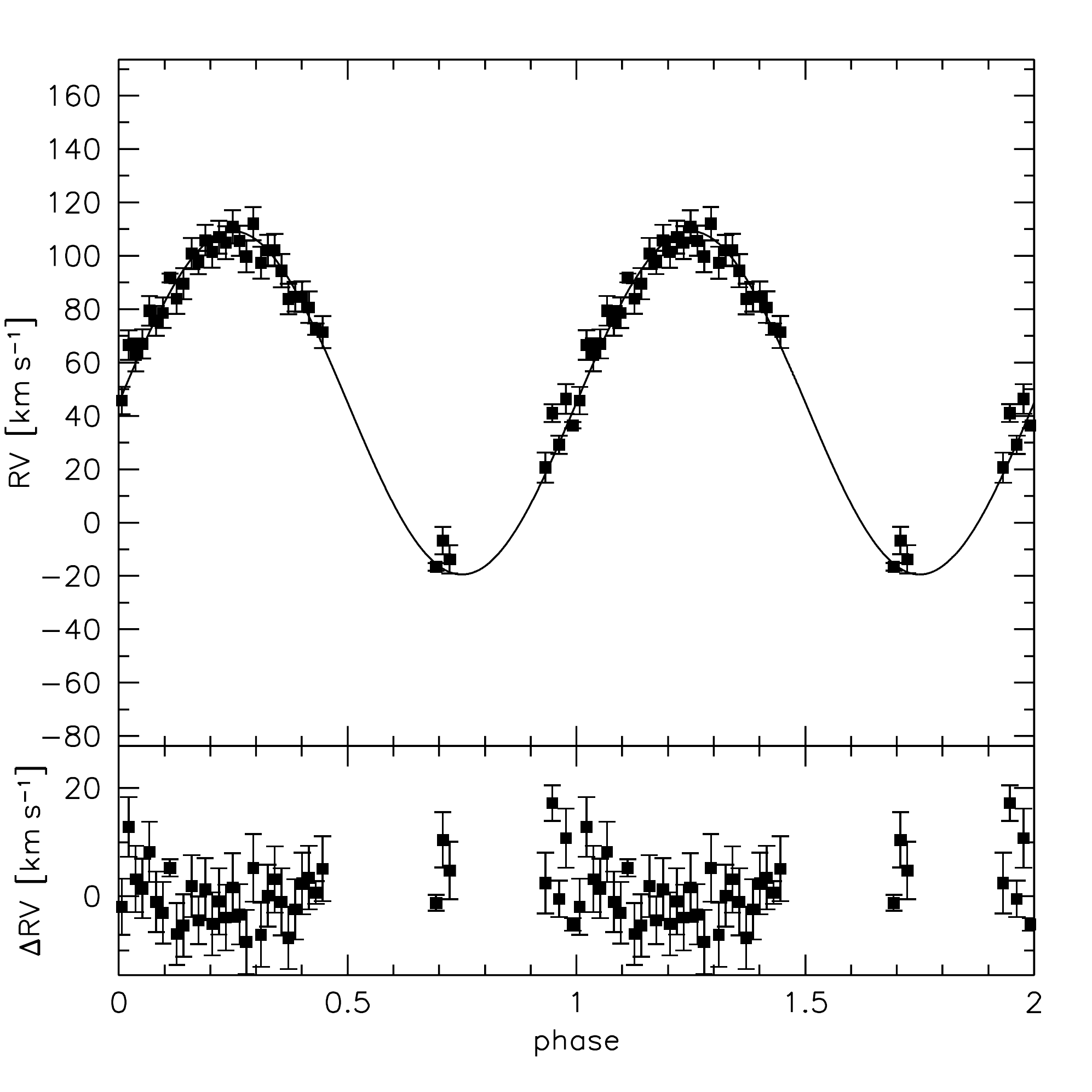}
\caption{Radial velocity curve of UVEX\,J032855.25+503529.8 derived from 38 spectra taken with WHT/ISIS. The three separated data points are the initial spectra. 
}
\label{fig:rv0328}
\end{figure}

\subsection{Results}\label{sec:results}
So far 5 targets from the UVEX survey and 45 targets from the GALEX survey were checked for RV variations. Three of the UVEX targets (1 He-sdO and 2 sdBs) were already identified by \citet{ver12} and two were colour-selected and identified as an sdb and as a cataclysmic variable, respectively. One of the 5 UVEX objects (UVEX\,J0328) was found to be a short period sdB binary (see Sec.~\ref{sec:uvex}).\\
From the 45 sdBs of the GALEX sample only one system (HS\,1741+2133) showed a significant shift on short timescales (see Sec.~\ref{sec:hs1741}).

\begin{figure}[t]
\centering
 \includegraphics[width=0.5\textwidth, angle=270]{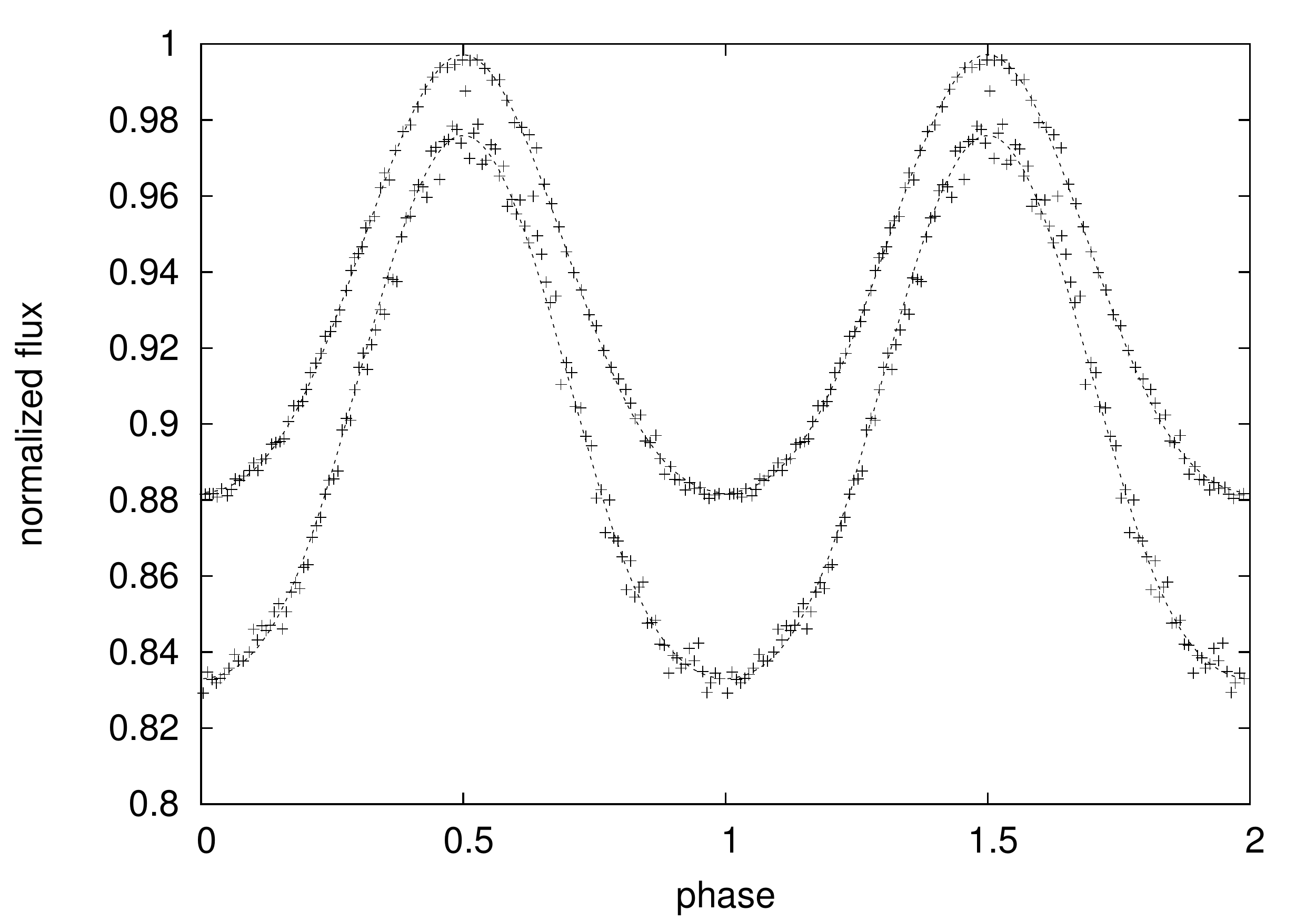}
\caption{Phase-folded and binned $R$ band (lower curve) and $g^{\prime}$ band (upper curve) lightcurve of UVEX\,J032855.25+503529.8 obtained with the pt5m telescope and the McDonald 1.0\,m telescope, respectively. The dotted line corresponds to the computed lightcurve for a $M_1$=0.49\,M$_\odot$ sdB primary and a $M_2$=0.12\,M$_\odot$ main sequence companion seen under an inclination angle of 54$^\circ$.}
\label{fig:lc0328}
\end{figure}

\subsubsection{UVEX\,J032855.25+503529.8}\label{sec:uvex}
UVEX\,J0328 ($g^{\prime}$=14.20\,mag) was identified by \citet{ver12} as a single-lined sdB ($T_{\rm eff}$=28\,500\,K, log\,$g$=5.5). A comparison between UVEX, IPHAS, UKIDSS, 2MASS and WISE photometry with a computed sdB model spectrum with $T_{\rm eff}$=30\,000\,K, log\,$g$=5.5 and $E$($B$--$V$)=0.4 showed that the photometry of UVEX\,J0328 can be fully explained by a reddened single sdB \citep{ver13}.\\
We observed UVEX\,J0328 again with the WHT using the ISIS spectrograph and identified the system to be RV variable within 10\,min and re-observed it in the same night for another 1.5\,h with an exposure time of 2\,min per spectrum. We took a total of 38 spectra. Fig.~\ref{fig:rv0328} shows the RV curve of the system. We could measure the RV curve in a single night and found a period $P$=0.1\,days and a RV semi-amplitude $K_1$=64\,km\,s$^{-1}$. Assuming that the primary has a canonical sdB mass of $M_1$=0.47\,M$_\odot$, a minimum companion mass of $M_2>$0.1\,M$_\odot$ is derived.\\
Variability was detected in a $R$ band lightcurve obtained with the 35\,cm teles\-cope mounted on the roof of the Huygensgebouw at the Radboud University \mbox{Nijmegen}. Therefore, additional lightcurves were obtained with the robotic 50\,cm pt5m telescope mounted on the roof of the William Herschel Telescope in the $R$ band as well as with the McDonald 1.0\,m telescope from the Las Cumbres Observatory Global Telescope Network \citep{bro13} with the $g^{\prime}$ filter. From the lightcurves of the three teles\-copes a period of $P$=0.11017\,days was derived which is consistent with the period derived from the RV curve. Fig.~\ref{fig:lc0328} shows the phased lightcurve. Clearly visible is a strong reflection effect in UVEX\,J0328. However, no eclipses are detected.\\ 
A first analysis of the pt5m and the McDonald 1.0\,m telescope lightcurves was done using the MORO code which is described in \citet{dre95}. From the mass function we estimated the mass ratio to be $M_2$/$M_1$=$q$=0.25 which was kept fixed in the analysis. The analysis of the lightcurve gave masses of $M_1$=0.49\,$\pm$\,0.05\,M$_\odot$ for the sdB and $M_2$=0.12\,$\pm$\,0.01\,M$_\odot$ for the companion. We can already state that UVEX\,J0328 is a new bright short period non-eclipsing sdB with M-dwarf companion (see Tab.~\ref{tab:par}).

\begin{table*}[t]
\begin{center}
 \caption{Derived parameter for the discovered sdb binaries}
  \begin{tabular}{ccccccc}
  \hline
   Object    & Period                    & $\gamma$              &  $K_1$                    &  $f$(M)        &    Companion \\
             &  [days]                   & [km\,s$^{-1}$]          &  [km\,s$^{-1}$]         &  [M$_\odot$]  &  \\
              \hline\hline
    HS\,1741     &  0.20\,$\pm$\,0.01      &  --112.8\,$\pm$\,2.7   &   157.0\,$\pm$\,3.4   & 0.08       &   WD \\
    UVEX\,J0328  &  0.11017\,$\pm$\,0.00011   &      44.9\,$\pm$\,0.7           &   64.0\,$\pm$\,1.5    &  0.0029      &  M-dwarf \\
    \hline
\end{tabular}
\label{tab:par}
\end{center}
\end{table*}

\begin{figure}[t]
\centering
 \includegraphics[width=0.5\textwidth]{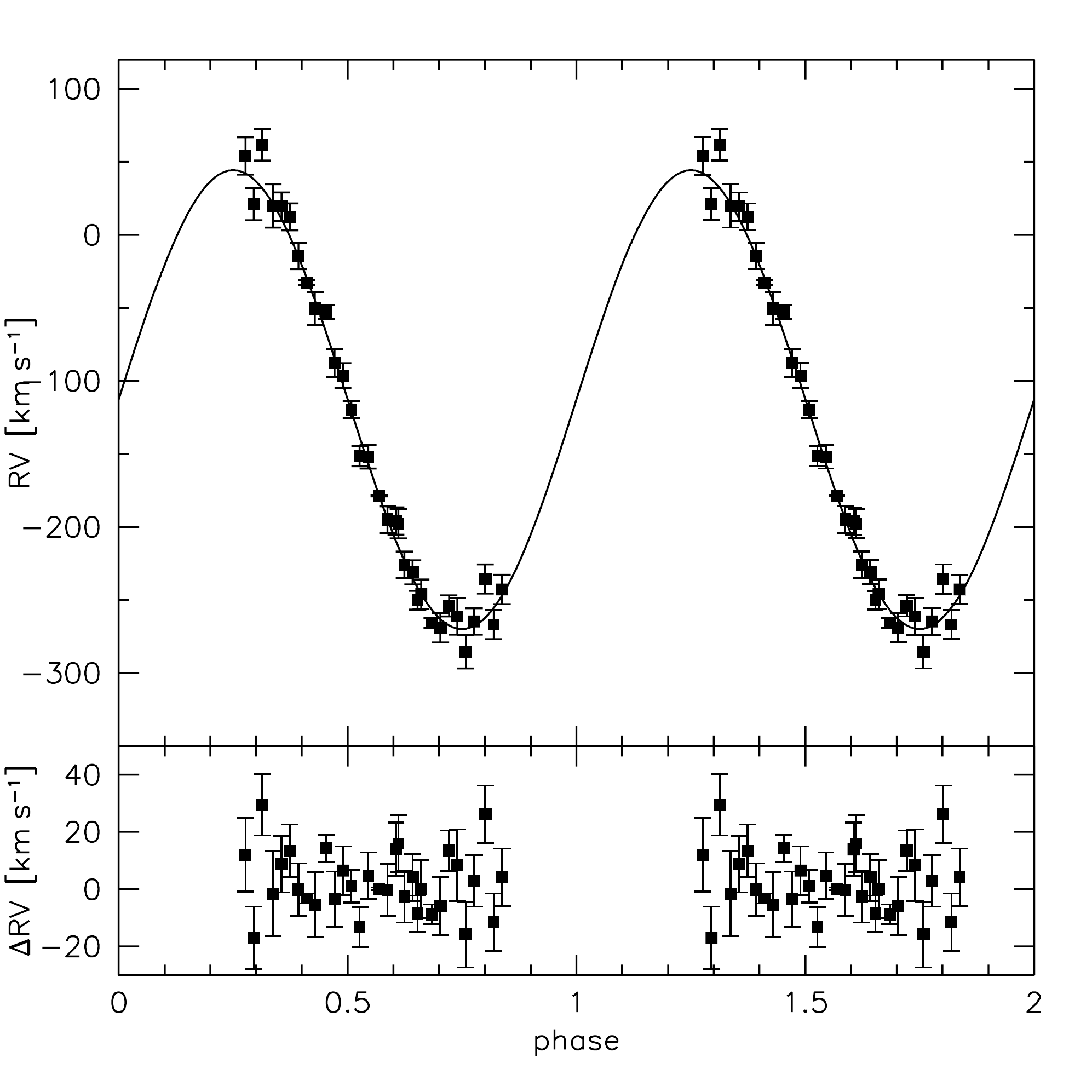}
\caption{Radial velocity curve of HS\,1741+2133 derived from 33 spectra taken with INT/IDS.}
\label{fig:rv1743}
\end{figure}

\subsubsection{HS\,1741+2133}\label{sec:hs1741}
HS\,1741 ($B$=13.63\,mag) was identified as an sdb candidate in the Hamburg Quasar Survey. \citet{ede03} derived parameters for the system and identified it as sdB with $T_{\rm eff}$=35\,600\,K and log\,$g$=5.3. \citet{dre02} analysed 250 photometric data points in white light obtained at the 1.2\,m telescope at Calar Alto Observatory. They found no photometric variability larger than 6\,mmag.\\
We observed HS\,1741 again with the INT using the IDS spectrograph and found a RV variability from three spectra and re-observed the object for another 2.5\,hours with an exposure time of 5\,min in the following night. These data are sufficient to determine the radial velocity curve (see Fig.~\ref{fig:rv1743}) and the mass function. We derived a period of $P$=0.20\,$\pm$\,0.01\,d and a RV semi-amplitude $K_1$=157.0\,$\pm$\,3.4\,km\,s$^{-1}$. As the lightcurve shows no variation we can only derive a lower limit for the companion mass. Assuming an sdb mass of 0.47\,M$_\odot$ we determined a minimum companion mass of $M_2>$0.39\,M$_\odot$. Due to the short period a reflection effect from a cool main sequence star should be visible if the companion is a low mass non-degenerated star. The lack of such a variation in the lightcurve leads to the conclusion that the companion most likely is a white dwarf (see Tab.~\ref{tab:par}). 

\subsection{Conclusion \& Outlook}
We have presented first results of an ongoing new campaign which aims at finding short period sdB/sdO binaries ($P<$0.5\,days depending on the companion mass) at low Galactic latitudes (--30$^\circ<b<$+30$^\circ$). We observed 50 targets and found 2 new compact sdB binaries in close orbits. The discoveries demonstrate that our strategy works and that we are able to detect RV variations on short timescales and measure RV curves with subsequent observations.The example of UVEX\,J0328 also showed that we are able to detect compact binaries with low mass companions and hence fairly low semi-amplitudes. \\
During future observing runs we will observe more systems and compare our findings with the known sdB binary population at high Galactic latitudes. As a side effect, we will also discover other interesting systems, e.g. cataclysmic variables, DB white dwarfs or DA+M-dwarf binaries in the UVEX sample.

\acknowledgements 
TK acknowledges support by the Netherlands Research School for Astronomy (NOVA). VS was supported by Deutsches Zentrum f\"ur Luft- und Raumfahrt DLR under grant 50 OR 1110.
CH und EZ were supported by Deutsche Forschungsgemeinschaft under grant He1356/62-1 and He1356/45-2, respectively. \\
Some results presented in this paper are based on observations made with the William Herschel Telescope and the Isaac Newton Telescope operated on the island of La Palma by the Isaac Newton Group in the Spanish Observatorio del Roque de los Muchachos of the Institutio de Astrofisica de Canarias. This paper uses observations obtained with facilities of the Las Cumbres Observatory Global Telescope.

\bibliography{author}
\bibliographystyle{asp2010}

\end{document}